\documentclass[prb,showpacs,twocolumn]{revtex4}
\usepackage{epsfig}
\usepackage{amsmath}
\begin{document}
\bibliographystyle{revtex}
\title{Charge transfer through molecular junctions within Redfield theory: \\
subtleties and pitfalls}
\author{Lars Kecke}
\affiliation{Institut f\"ur theoretische Physik, Universit\"at Ulm,
Albert-Einstein-Allee 11, D-89069 Ulm, Germany}
\author{Joachim Ankerhold}
\affiliation{Institut f\"ur theoretische Physik, Universit\"at Ulm,
Albert-Einstein-Allee 11, D-89069 Ulm, Germany}

\date{\today}

\begin{abstract}
Charge transfer through nanoscale junctions connecting metallic leads with quantum dots or single molecules is often described within an open system formulation in terms of Redfield theory.
Under non-equilibrium conditions, the usually invoked rotating wave approximation is not justified which may lead to unphysical
steady state solutions with e.g.\ negative populations. In this work we explore subtleties and constraints of the approach and thus clarify its applicability in numerical calculations. General findings are illustrated for an analytically solvable case of a molecule with two electronic states.
\end{abstract}


\maketitle

\section{Introduction}

Electronic transport through nanoscale junctions has been in the focus of
intensive research for several years now, particularly in the context of tailored quantum dot structures
and in the context of molecular electronics\cite{reviews,cuniberti,scheer}.
Theoretical descriptions are challenging
as they have to capture the complex non-equilibrium many-body physics due to the interaction of electronic
and phonon states on the dot/molecule with those in the leads.
Several strategies have been developed in the past, each with its strengths and each with its limitations.
 Ab initio approaches start from the chemical microstructure of the dot/molecule-leads compound but have difficulties to
 include e.g.\ strong correlation effects, strong non-equilibrium, and inelastic tunneling processes.
Efficient formulations to describe these many-body phenomena work with model Hamiltonians which have been
very successfully applied for decades in mesoscopic physics. Here, theoretical studies of charge transfer
have been put forward  either within the well-established framework of single-particle non-equilibrium Greens functions\cite{meirwingreen,galperin,greens} or within the context of open quantum systems \cite{weiss,breuer,gurvitz,millis}.
The former ones consider  the dot/molecule as a tunneling barrier with energy-dependent
transmission, while for the latter ones the leads constitute fermionic heat baths which must be
properly eliminated to arrive at a reduced description of the dot/molecule degrees of freedom \cite{semigroups,timm,grifoni,wegewijs,timm2,grifoni2}.

In both cases, approximate treatments are required to access the charge transfer process, thus concentrating on various domains in
parameter space such as, for example, the single charge sector for weak tunnel couplings including inelastic processes, on many body states on the intermediate aggregate, or on elastic scattering in the coherent regime.
However, whenever the fundamental physics changes by tuning external parameters (bias voltage, gate voltage, coupling, temperature etc.),
one typically also has to switch to a complementary approach to avoid inconsistencies or even unphysical results \cite{pride}.

In the context of a reduced description, a powerful tool are master equations (ME) based on a perturbative
treatment of the interaction between the relevant agent and its surrounding. In case of bosonic heat baths
this leads to the famous Redfield equation \cite{redfield,breuer} or, with additional constraints imposed, to Lindblad theory \cite{lindblad,semigroups}.
Things become more intricate though when the environment drives the agent out of equilibrium and towards a steady state as in the case of voltage-biased tunnel junctions. Namely, under these circumstances the usually invoked rotating wave approximation (RWA) which decouples the dynamics of diagonal elements of the reduced density (populations) from those of the off-diagonal ones (coherences)
is not justified. Mathematically, the existence of complete positive densities is then no longer guaranteed. Apart from the failure of this formal concept, the question arises to what extent or under which conditions physical acceptable steady states, i.e.\ those with positive populations, are actually supported by  MEs in practical applications.

To shed some light on this issue, is the intention of the present work. Some of these and related questions have been addressed already previously \cite{timm,wegewijs,grifoni2}, but to our knowledge no consistent deeper analysis has been given yet.
As we are interested in generic properties, we consider a strongly simplified
set-up compared to realistic junctions, namely, a single dot/molecule with discrete electronic states interacting with broadband metallic leads (Sec.~\ref{system}). Integrating out the lead degrees of freedom brings us to the conventional Redfield equation with the
Redfield tensor as its central ingredient (Sec.~\ref{redfield}).
In steady state the underlying Markov approximation always applies so that the proper
expansion parameter for the Born series is not immediately obvious. We formulate constraints that the Redfield dissipator has to obey for physical steady states to exist as stationary solutions of the Redfield equations.
It turns out that already the zero-voltage situation must be treated with care since the density operator of the compound may significantly differ from the product state of its bare constituents.
General findings are illustrated by considering a molecule with only two electronic levels which allows for a complete analytical solution, Sec.~\ref{example}. In the linear response regime contact is made with the non-equilibrium Greens function formulation (Sec.~\ref{perturbation}).
Conclusions summarize our results and specify under which conditions a Redfield approach is applicable for charge transfer (Sec.~\ref{conclusions}).

\section{System and Redfield formulation}\label{system}
We consider a minimal model consisting of  a quantum system with Hamiltonian
\begin{equation}
H_0=\sum_a e_a|a\rangle\langle a|
\end{equation}
which is positioned between two fermionic leads $l=L,R$ with
\begin{equation}
H_B=\sum_\alpha \epsilon_\alpha c^\dagger_\alpha c_\alpha\, ,
\end{equation}
where the sum is over a multi-index $\alpha=(k,l)$. In the sequel, we will keep this notation and  use
Greek letters for lead states and Latin letters for dot states.
As usual, electronic states in the leads are  Fermi distributed
$$
\langle c^\dagger_\alpha c_\alpha\rangle=\frac{1}{\exp(\beta\epsilon_\alpha)+1}
=f(\epsilon_\alpha)
$$
at inverse temperature $\beta$ and chemical potential $V_l$ so that
$\epsilon_\alpha=\epsilon_k-V_l$ with the single particle energy $\epsilon_k=\hbar^2k^2/2 m$.

The tunneling of electrons between leads and dot is described by
\begin{equation}
H_I=\sum_{\alpha} \gamma_\alpha(c^\dagger_\alpha\Phi_\alpha+\Phi_\alpha^\dagger c_\alpha)
\, .\label{HI}
\end{equation}
Here, the operator $\Phi_\alpha$ removes an electron from the dot to state $\alpha$.

At vanishing bias voltage the full equilibrium density operator can be written as
$W=\exp[-\beta (H_0+H_B+H_I)]/Z$, where the partition function is given by
$Z={\rm Tr}\{\exp[-\beta (H_0+H_B+H_I)]\}$.
The reduced density operator of the dot is given by tracing out the leads, i.e.,
$\rho={\rm Tr}_B\,W$. In the regime of weak lead-dot coupling, one could then assume that
to leading order $W\approx W_{\rm prod}=\rho_0\otimes W_B\sim {\rm e}^{-\beta H_0}\, {\rm e}^{-\beta H_B}$.
Such an ansatz is, of course, only justified as long as higher order terms are sufficiently small. However, this is not always guaranteed.
Namely, exploiting the Baker-Hausdorff formula one has for the next order contributions (see Ref.~\onlinecite{BCH})
\begin{eqnarray}\label{equi}
W&\approx&
W_{\rm prod}\cdot\exp(-\beta H_I)\cdot\exp\left(\frac{\beta^2}{2}[H_0+H_B,H_I]\right)\nonumber\\
&&\cdot\exp\left(\frac{\beta^3}{12}[[H_0+H_B,H_I],H_I]\right)\nonumber\\
&&\cdot\exp\left(\frac{\beta^3}{12}[[H_I,H_0+H_B],H_0+H_B]\right)\, .
\end{eqnarray}
Due to the fact that $H_B\propto c^\dagger c$ and $H_I\propto \Gamma\,  (c+c^\dagger)$ with typical coupling parameter $\Gamma$, the double commutator $[[H_B,H_I],H_I]$ is proportional to $\Gamma^2$ times the identity in the lead-subspace. Taking the trace may thus yield an ill-defined reduced dot operator depending on the operators $\Phi_\alpha$ but no matter how small the coupling is.
The physical reason for that is easy to understand: Hybridization between lead and dot states immediately occurs for $\Gamma\neq 0$ such that for e.g.\ Fermi levels in the leads lying above the lowest unoccupied level of the dot (LUMO), the ground state of the reduced lead-dot compound is orthogonal to that of the bare dot. Accordingly, microscopic approaches such as density functional theory (DFT) calculate the equilibrium state of the coupled dot or molecule by including part of the leads (the so-called super-molecule)\cite{scheer}.

At a finite bias voltage, charge transfer sets in and we are interested in the corresponding steady state. This is approached during the time evolution of the compound starting
from an initial state, e.g.\ an uncoupled equilibrium state of system and leads.
One way to do so, is to derive an equation of motion for the reduced density of the dot for weak tunnel couplings.
Here ''weak'' is often simply understood as a small coupling parameter $\Gamma$. One purpose of this work is to
reveal more precisely in which sense such a perturbative treatment actually applies.
We thus proceed with a conventional Born-Markov approximation\cite{weiss,breuer,millis} and start by expanding
the Liouville-von Neumann equation of the full system $-i\hbar\frac{{\rm d}W}{{\rm d}t}=[W,H]$
up to second order in $H_I$.
As a result one obtains  the well-known equation \cite{breuer,millis}
\begin{eqnarray}\label{master1}
\frac{d\rho}{dt}&=&\frac{i}{\hbar}[\rho,H_0]-
\frac{1}{\hbar^2}\int_0^\infty {\rm d}\tau
{\rm Tr}_B \{[H_I,[H_{I}(-\tau),W(t)]]\}\nonumber\\
&=&\frac{i}{\hbar}[\rho(t),H_0]+R\rho(t)\, ,
\end{eqnarray}
where the Redfield tensor $R$ accounts for the impact of the interaction
Hamiltonian $H_I$. The time dependence in the integral follows from the interaction picture
$H_I(-\tau)=U_0 H_I U_0^\dagger=
\exp[-\frac{i\tau}{\hbar}(H_0+H_B)]H_I\exp[\frac{i\tau}{\hbar}(H_0+H_B)]$. The central assumption in (\ref{master1}) are macroscopic reservoirs such that $W(t)=\rho(t) \otimes\exp(-\beta H_B)$.

Steady states are now inferred as the stationary solutions to the ME. The Markov approximation in (\ref{master1}), i.e.\ a sufficiently short-ranged reservoir correlation function, is exact in this case. A subtlety arises, however, namely, whether one searches for solutions in the interaction or in the Schr\"odinger picture \cite{grifoni2}.
The first one is determined by $R\rho_{st}=0$, while the second one follows from
$R\rho_{S, st}=-\frac{i}{\hbar}[\rho_{S, st},H_0]$. It is well-known though that the latter ones never carry any coherences, i.e., finite off-diagonal elements of $\rho_{S, st}$,  in a consistent second
order perturbation theory and as long as the energy spacings of the states $|a\rangle$ are larger than the tunnel coupling
constant $\Gamma=\sum_k\gamma_k^2\frac{\partial k}{\partial\epsilon}$ (see Ref. \onlinecite{wegewijs}). The argument is simple:
During the time evolution off-diagonal elements $\langle a|\rho_{S}(t)|b\rangle$ oscillate with frequency $\omega_{ab}=(\epsilon_a-\epsilon_b)/\hbar$ and for large times are thus washed out. This is not the case in the interaction picture where effectively one works in a rotating frame and may thus be able to keep reservoir induced coherences between dot/molecule states. 

Once the steady state is at hand, the most interesting quantity one can calculate  is the current through lead $l$:
\begin{eqnarray}
\langle I_l\rangle &=&e\lim_{t\to \infty}\langle \dot{N}_l\rangle =
e\lim_{t\to \infty} {\rm Tr}\{N_l\dot{W(t)}\}\nonumber\\
&=&-\frac{ie}{\hbar}{\rm Tr}\{N_l[H_0,\rho_{\rm st}]\}\nonumber\\
&&\hspace{0.7cm}-\frac{e}{\hbar^2}\int_0^\infty{\rm d}\tau\,{\rm Tr}\{N_l[H_I,[H_I(-\tau),W_{\rm st}]]\}\nonumber\\
&=&-\frac{e}{\hbar^2}\int_0^\infty{\rm d}\tau\,{\rm Tr}
\{\rho_{\rm st}\langle[J_l,H_I(-\tau)]\rangle_B\}\label{ieq}
\end{eqnarray}
with the lead $l$ number operator $N_l=\sum_\alpha \delta_{\alpha, l}
c^\dagger_\alpha c_\alpha$ ($\delta_{\alpha, l}$ is the Kronecker symbol which is only non-vanishing if the lead index in $\alpha$ is equal to $l$) and the current operator
$J_l=[N_l,H_I]=\sum_\alpha \delta_{\alpha,l}\gamma_\alpha
(\Phi_\alpha c^\dagger_\alpha-\Phi_\alpha^\dagger c_\alpha)$. In turns out that in state representation the current is determined by coherences of the steady state (see below).

Unfortunately, the steady state operator $\rho_{\rm st}$ obtained from (\ref{master1}) does not have to be positive though, as we show now.
For this purpose, we consider the fermion propagators of the leads
\begin{equation}
\kappa_\alpha(\tau):=\langle c_\alpha^\dagger(\tau)c_\alpha(0)\rangle_B
=e^{\frac{i}{\hbar}\epsilon_\alpha t}f(\epsilon_\alpha-V_l)\,
\end{equation}
and
\begin{eqnarray}
\kappa^*_\alpha(\tau)&:=&\langle c_\alpha^\dagger(-\tau)c_\alpha(0)\rangle_B
=\exp(-\frac{i}{\hbar}\epsilon_\alpha\tau)f(\epsilon_\alpha-V_l),\nonumber\\
\bar{\kappa}_\alpha(\tau)&:=&\langle c_\alpha(\tau)c^\dagger_\alpha(0)\rangle_B
=\exp(-\frac{i}{\hbar}\epsilon_\alpha\tau)f(V_l-\epsilon_\alpha),\nonumber\\
\bar{\kappa}^*_\alpha(\tau)&:=&\langle c_\alpha(-\tau)c^\dagger_\alpha(0)\rangle_B
=\exp(\frac{i}{\hbar}\epsilon_\alpha\tau)f(V_l-\epsilon_\alpha)\,.
\end{eqnarray}
Now, after cyclic rearrangement of the $c_\alpha$ and $c^\dagger_\alpha$ under the trace, one arrives at
\begin{eqnarray}\label{redfield1}
R\rho&=&-\int_0^\infty{\rm d}\tau\sum_{\alpha}\frac{\gamma_\alpha^2}{\hbar^2}
\left[\kappa_\alpha\Phi_\alpha \Phi_\alpha^\dagger (-\tau)\rho
+\bar{\kappa}_\alpha\Phi^\dagger \Phi_\alpha(-\tau) \rho
\right.\nonumber\\
&+&\kappa_\alpha^*\rho \Phi_\alpha (-\tau)\Phi_\alpha^\dagger+\bar{\kappa}_\alpha^*\rho \Phi_\alpha^\dagger (-\tau)\Phi_\alpha
-\bar{\kappa}_\alpha^* \Phi\rho \Phi_\alpha^\dagger (-\tau)\nonumber\\
&-&\left.\kappa_\alpha^* \Phi_\alpha^\dagger\rho \Phi_\alpha(-\tau)
-\kappa_\alpha \Phi_\alpha (-\tau)\rho\Phi_\alpha^\dagger
-\bar{\kappa}_\alpha \Phi_\alpha^\dagger (-\tau)\rho\Phi_\alpha \right],\nonumber\\
\end{eqnarray}
where in state representation $\Phi_{\alpha,ab}(-\tau)\equiv\langle a|\Phi_\alpha(-\tau)|b\rangle=
\Phi_{\alpha,ab}\exp( i\omega_{ab}\tau)$ with $\hbar\omega_{ab}=(e_a-e_b)/\hbar$.

Now, according to Lindblad mathematically complete positivity of $\rho_{\rm st}$ is only guaranteed if
the dissipator $R$ is of the form \cite{lindblad,semigroups}
$$
R_{L}=\sum_j\left\{[L_j\rho,L_j^\dagger]+[L_j,\rho L_j^\dagger]\right\}\, .
$$
In the above (\ref{redfield1}), one then must have
\begin{eqnarray}
\kappa_\alpha\Phi_\alpha\Phi^\dagger_\alpha(-\tau)
&+&\bar\kappa_\alpha\Phi_\alpha^\dagger\Phi_\alpha(-\tau)=\nonumber\\
&&\kappa_\alpha^*\Phi_\alpha(-\tau)\Phi^\dagger_\alpha
+\bar\kappa_\alpha^*\Phi_\alpha^\dagger(-\tau)\Phi_\alpha\nonumber\\
&\equiv & L^\dagger L\,.
\end{eqnarray}
This can be archived  if (i) one applies in addition a secular approximation \cite{breuer} (also called rotating wave approximation, RWA)
so that the $\exp(-i\epsilon_\alpha\tau)$ in the $\kappa$ cancels the $\exp(-i\hbar\omega_{ab}\tau)$ in the $\Phi(-\tau)$,
or if (ii) one drops the $\tau$-integration (this leads to the singular coupling approximation \cite{breuer}) or if (iii) one 'symmetrizes' the range of integration $\int_0^\infty d\tau\to (1/2) \int_{-\infty}^\infty d\tau$. However, under non-equilibrium conditions (i) may not be justified (see e.g.~\cite{pekola}), (ii) is in general not realized, and (iii) is an {\em ad hoc} procedure.
This is in contrast for e.g.\ a bosonic heat bath interacting with an autonomous  system, where the asymptotic state of the reduced system is the bare thermal equilibrium $\sim{\rm e}^{-\beta H_0}$ so that coherences (off-diagonal elements) die out during the time evolution. The RWA or related approximations are then applicable under much milder conditions.
Here, if none of these reductions work, one may wonder if (or under which conditions) at least positivity (all populations $\langle a|\rho_{\rm st}|a\rangle\geq 0$) survives so that stationary solutions of (\ref{redfield1}) yield physical steady states. This will be the central subject of the remainder of this work.

\section{The Redfield dissipator}\label{redfield}
In the following, in state representation we think of $R=R_{(ab,cd)}$ in (\ref{redfield1}) as a matrix with 'columns' $(ab)$ and 'rows' $(cd)$ which then can be written in the form
\begin{eqnarray}
R_{ab,cd}&=&\delta_{bd}\sum_{c',\alpha}
(\Phi_{\alpha,ac'}^\dagger\Phi_{\alpha,c'c}\Sigma^{*{\rm out}}_{\alpha,c'c}
+\Phi_{\alpha,ac'}\Phi_{\alpha,c'c}^\dagger\Sigma^{*{\rm in}}_{\alpha,c'c})\nonumber\\
&&+\delta_{ac}\sum_{c',\alpha}
(\Phi_{\alpha,dc'}^\dagger\Phi_{\alpha,c'b}\Sigma^{\rm out}_{\alpha,c'd}
+\Phi_{\alpha,dc'}\Phi_{\alpha,c'b}^\dagger\Sigma^{\rm in}_{\alpha,c'd})\nonumber\\
&&-\sum_\alpha
[\Phi_{\alpha,db}^\dagger\Phi_{\alpha,ac}
\left(\Sigma^{\rm out}_{\alpha,bd}+\Sigma^{*{\rm out}}_{\alpha,ac}\right)\nonumber\\
&&\hspace{1cm}+\Phi_{\alpha,db}\Phi_{\alpha,ac}^\dagger
\left(\Sigma^{\rm in}_{\alpha,bd}+\Sigma^{*{\rm in}}_{\alpha,ac}\right)]\, .\label{Rab}
\end{eqnarray}
Here we introduced the self energy
\begin{equation}
\Sigma^{\rm in}_{\alpha,ab}=\gamma_\alpha\int_0^\infty {\rm d}\tau
\kappa_\alpha(-\tau)\exp(i\omega_{ab}\tau)
\end{equation}
which is identical to the  Keldysh inscattering function
[\onlinecite{datta}] if the transition from state $b$ to $a$ moves an electron
from lead to dot, respectively. The corresponding outscattering function reads
$
\Sigma^{\rm out}_\alpha=\gamma_\alpha\int_0^\infty {\rm d}\tau
\bar{\kappa}_\alpha(-\tau)\exp(i\omega_{ab}\tau)
$.
Thus, the time evolution of the populations
$\rho_{aa}$ is governed by
\begin{eqnarray}
R_{aa,cd}&=&\delta_{ad}
\sum_{\alpha,c'}\Phi_{\alpha,ac'}^\dagger\Phi_{c'c}
\Sigma_{\alpha,c'c}^{*{\rm out}}
+\delta_{ac}\sum_{\alpha,c'}\Phi_{\alpha,dc'}^\dagger\Phi_{\alpha,c'a}
\Sigma_{c'd}^{\rm out}\nonumber\\
&&-\sum_\alpha\Phi_{\alpha,da}^\dagger\Phi_{ac}
(\Sigma_{\alpha,ad}^{\rm out}+\Sigma_{\alpha,ac}^{*{\rm out}})
+\{{\rm in \leftrightarrow out}\}\, .
\end{eqnarray}
This matrix is in general complex-valued and couples not only populations, but also populations and coherences.

Now, due norm conservation the row trace of $R$ is zero ({\em Property 1}), i.e.,
\begin{eqnarray}
\forall_{cd}\sum_{\alpha,a}R_{aa,cd}&=&\sum_{c'}\Phi_{\alpha,dc'}^\dagger\Phi_{\alpha,c'c}
(\Sigma_{\alpha,c'c}^{*{\rm out}}+\Sigma_{\alpha,c'd}^{\rm out})
\nonumber\\
&&-\sum_{\alpha,a}\Phi_{\alpha,da}^\dagger\Phi_{\alpha,ac}
(\Sigma_{\alpha,ad}^{\rm out}+\Sigma_{\alpha,ac}^{*{\rm out}})\nonumber\\
&&\hspace{1cm}+\{{\rm in \leftrightarrow out}\}=0\,.\label{rsum}
\end{eqnarray}
This in turn means that the rank of $R$ is not full so that $R\rho_{\rm st}=0$ always has a non-vanishing solution.

In particular the stationary current through the dot (\ref{ieq}) is thus given by
\begin{eqnarray}
\langle I_l\rangle
&=&\frac{e}{\hbar^2}\int_0^\infty{\rm d}\tau {\rm Tr}_s\{
\rho_{\rm st}\langle[J,H_I(-\tau)]\rangle_l\}\nonumber\\
&=&\frac{e}{\hbar^2}\sum_{\alpha,abc}\delta_{\alpha,l}\rho_{{\rm st},ab}
(\Phi_{\alpha,bc}\Phi_{\alpha,ca}^\dagger\Sigma_{\alpha,ba}^{\rm in}\nonumber\\
&&\hspace{1.3cm}-\Phi_{\alpha,bc}^\dagger\Phi_{\alpha,ca}\Sigma_{\alpha,ba}^{\rm out})\, .
\label{ieq2}
\end{eqnarray}

\subsection{Self energy}
To further elucidate the properties of the Redfield dissipator, we explicitly calculate the self-energies $\Sigma$. Accordingly, the wide band limit is taken and coupling constants are assumed to be independent of the fermion state $\gamma_\alpha=\gamma_l$ with
bands of constant density of states $D_l$ in the leads. As a consequence, one has
\begin{eqnarray}
\Sigma^{\rm in}_{l,ab}&=&\sum_\alpha \delta_{\alpha,l}\Sigma_{\alpha,ab}^{\rm in}\nonumber\\
&=&\int_0^\infty{\rm d}\tau\sum_\alpha\delta_{\alpha,l}\gamma_\alpha^2
e^{\frac{i}{\hbar}\epsilon_a\tau}\langle c^\dagger_k(-\tau)c_k(0)\rangle_l
e^{-\frac{i}{\hbar}\epsilon_b\tau}\nonumber\\
&=&\Gamma_l\int_0^\infty{\rm d}\tau e^{i(\omega_{ab}-V_l)\tau}
\int{\rm d}\epsilon e^{-\frac{i}{\hbar}\epsilon \tau}f(\epsilon)
\end{eqnarray}
with $\Gamma_l=D_l\gamma^2_l$. The second integral yields
\begin{equation}
\int {\rm d}\epsilon\  {\rm e}^{-\frac{i}{\hbar}\epsilon\tau}f(\epsilon)=
\frac{i\pi}{\beta\sinh\frac{\pi\tau}{\beta}}\, ,
\end{equation}
where $\tau$ includes a positive imaginary increment $i/\omega_c$.
Therefore, with the abbreviation $\omega_{l,ab}=\omega_{ab}-V_l/\hbar$ we have
\begin{eqnarray}\label{selfenergy}
\Sigma^{\rm in}_{l,ab}&=&\frac{i\pi\Gamma_l}{\hbar\beta}
\int_{i/\omega_c}^\infty{\rm d}\tau
\frac{{\rm e}^{i\omega_{l,ab}\tau}}{\sinh\frac{\pi\tau}{\hbar\beta}}\nonumber\\
&=& \frac{2 i\pi\Gamma_l}{\hbar\beta}
\int_{i/\omega_c}^\infty{\rm d}\tau {\rm e}^{i\omega_{l,ab}\tau}\sum_{n=0}^\infty {\rm e}^{-\nu_{n+1/2}\tau}\nonumber\\
&=&-i\Gamma_l\left[\gamma_E+\Psi
\left(\frac{1}{2}-i\frac{\omega_{l,ab}}{\nu_1}\right)+\frac{i\pi}{2}
+\ln\left(\frac{\nu_1}{\omega_c}\right)\right]\nonumber\\
&=:&\Sigma(\omega_{ab}-V_l)\,,
\end{eqnarray}
with Matsubara frequencies $\nu_n=2\pi n/\hbar\beta$, Euler's constant $\gamma_E$, and in the limit where $\omega_c$ by far exceeds intrinsic energy scales of the dot. Physically, $\hbar\omega_c$ corresponds to a cut-off energy in the leads and must thus assumed to be large in order to be consistent with the Markov-approximation and the broadband limit.
For the real part one regains the Fermi distribution $\Sigma_l^{{\rm in}'}=\pi\Gamma_l f(\hbar\omega_{l,ab})$.
 The
imaginary part $\Sigma^{{\rm in}''}(\omega)$ is symmetric in $\omega$ and exhibits a logarithmic dependence on the cut-off $\omega_c$. Due to  $\Sigma_{ab}^{\rm out}(\omega)=\Sigma_{ab}^{in}(-\omega)$ in- and out-self energies contain identical
 imaginary parts $\Sigma^{{\rm in}''}$. Note that a symmetrization of the $\tau$-integration in (\ref{redfield1}) as discussed above,
 would only yield real-valued self-energies.  For a bosonic heat bath the corresponding imaginary part is known as reservoir induced Lamb-shift which differs from the one obtained for the fermionic self energy only by the additional term of  $1/2$ in the argument of the $\Psi$-function
\cite{breuer,vera}.

\subsection{Steady state}
We now search for a stationary solution $R\rho_{\rm st}=0$ for the density matrix in
interaction representation, which, thanks to {\em Property 1} always exists
and has a trace of 1. To be a physical density, $\rho_{\rm st}$  also has to be
positive definite. Equivalently, $|\rho_{ab}|^2\le\rho_{aa}\rho_{bb}$ which means that
all $\rho_{aa}\geq 0$. This in turn forces the row vectors of $R$ to be orthogonal on the subspace $\rho_{aa}>0$.
Accordingly ({\em Property 2}):  Stationary solutions of $R\rho_{\rm st}=0$ are {\em not} physical steady states if
\begin{equation}\label{prop2}
\ \ \exists_{ab}\ \forall_c\  R_{ab,cc}>0\, .
\end{equation}
In other words, rows of $R$ must contain at least one negative element for a physical steady state to exist.

In the secular approximation, i.e. by dropping all parts of $R$ which mix
coherences and populations, the existence of a {\em physical} solution $\rho_{\rm st}$ is always guaranteed. Then,  the conservation of
the total probability renders the rows of $R$ to be linear dependent so that each
column has a signature of the form $+----$ (the $i$th column has its plus at the $i$th position)
in the subspace of the $\rho_{aa}$, thus violating (\ref{prop2}).
For the full Redfield dissipator (\ref{Rab}), however, ''off-diagonal'' rows of $R$ do not need to
obey a certain signature and the existence of a physical acceptable solution depends on details of the problem under investigation. To analyze this analytically more carefully, we consider a two level system in the next section.

\section{Example: Two level system}\label{example}
The system consists of two electronic sites $|1\rangle$ and $|2\rangle$
with energies $e_{1/2}=\pm\delta/2$ and a hopping element $\Delta$ between
them. Site $|1\rangle$ couples to the left lead with a coupling strength
$\gamma_L$ and $|2\rangle$ couples to the right lead with a coupling strength
of $\gamma_R$. Diagonalizing the uncoupled Hamiltonian yields the singly
occupied energy eigenstates
\begin{eqnarray}
|+\rangle&=&\cos\frac{\phi}{2}\, |1\rangle+\sin\frac{\phi}{2}\, |2\rangle\nonumber\\
|-\rangle&=&-\sin\frac{\phi}{2}\, |1\rangle+\cos\frac{\phi}{2}\, |2\rangle
\end{eqnarray}
with eigenvalues $H_0|\pm\rangle=\pm\lambda|\pm\rangle$ where
$\lambda=\sqrt{\delta^2+\Delta^2}$ and the mixing angle is
$\tan(\phi/2)=\Delta/(\lambda+\delta)$.
Hopping to and from the dot (with the empty dot denoted with
$|0\rangle$ at energy $e_0$) is described by the Hamiltonian (\ref{HI})
with matrix elements
\begin{eqnarray}
\Phi_{L,0+}=\cos\frac{\phi}{2}&,&
\Phi_{R,0+}=\sin\frac{\phi}{2}\nonumber\\
\Phi_{L,0-}=-\sin\frac{\phi}{2}&,&
\Phi_{R,0-}=\cos\frac{\phi}{2}\,
\end{eqnarray}
and $\Phi_{l,aa}=\Phi_{l,+-}=0$.

\subsection{Redfield tensor}
Now, due to charge quantization, in the relevant density matrix
\begin{equation}
\rho=\left(\rho_{00},\rho_{++},\rho_{--},\rho_{+-},\rho_{-+}\right)\,,
\end{equation}
sectors with a different number of charges on the dot do not mix.
With the shorthand notation
\begin{equation}
\Phi^2_+:=\Phi_{+0}\Phi_{0+}^\dagger\ ,\
\Phi^2_-:=\Phi_{-0}\Phi_{0-}^\dagger=1-\Phi^2_+\ ,\
\Pi:=\Phi_{+0}\Phi_{0-}^\dagger\,,
\end{equation}
and using the fact that $\Sigma_{\pm0}=\Sigma^{\rm in}_{\pm0}$ and
$\Sigma_{0\pm}=\Sigma^{\rm out}_{0\pm}$
the Redfield matrix (\ref{Rab}) $R=R'+iR''$ reads
\begin{widetext}
\begin{eqnarray}
R'&=&\sum_l\nonumber\\
&&\left(
\begin{array}{ccccc}
2\Phi_+^2\Sigma'_{+0}+2\Phi_-^2\Sigma'_{-0}&
-2\Phi_+^2\Sigma'_{0+}&
-2\Phi_-^2\Sigma'_{0-}&
-\Pi(\Sigma'_{0-}+\Sigma'_{0+})&
-\Pi(\Sigma'_{0+}+\Sigma'_{0-})\\
-2\Phi_+^2\Sigma'_{+0}&2\Phi_+^2\Sigma'_{0+}&0&
\Pi\Sigma'_{0-}&\Pi\Sigma'_{0-}\\
-2\Phi_-^2\Sigma'_{-0}&0&2\Phi_-^2\Sigma'_{0-}&
\Pi\Sigma'_{0+}&\Pi\Sigma'_{0+}\\
-\Pi(\Sigma'_{-0}+\Sigma'_{+0})&
\Pi\Sigma'_{0+}& \Pi\Sigma'_{0-}&
\Phi_-^2\Sigma'_{0-}+\Phi_+^2\Sigma'_{0+}&0\\
-\Pi(\Sigma'_{+0}+\Sigma'_{-0})&
\Pi\Sigma'_{0+}& \Pi\Sigma'_{0-}&
0&\Phi_+^2\Sigma'_{0+}+\Phi_-^2\Sigma'_{0-}
\end{array}
\right)_l\nonumber\\
R''&=&\sum_l
\left(
\begin{array}{ccccc}
0&0&0&
\Pi(\Sigma''_{0+}-\Sigma''_{0-})&
\Pi(\Sigma''_{0-}-\Sigma''_{0+})\\
0&0&0&
\Pi\Sigma''_{0-}&-\Pi\Sigma''_{0-}\\
0&0&0&
-\Pi\Sigma''_{0+}&\Pi\Sigma''_{0+}\\
\Pi(\Sigma''_{+0}-\Sigma''_{-0})&\Pi\Sigma''_{0+}&-\Pi\Sigma''_{0-}&
-\omega_{\rm LS}&0\\
\Pi(\Sigma''_{-0}-\Sigma''_{+0})&-\Pi\Sigma''_{0+}&\Pi\Sigma''_{0-}&
0&\omega_{\rm LS}
\end{array}
\right)_l\, ,
\end{eqnarray}
\end{widetext}
where the lead index $l$ applies to all entries of the matrix. In the diagonal of $R''$, the imaginary part of the self energy gives rise to a
Lamb shift
\begin{equation}
\omega_{\rm LS}=\sum_l(\Phi_{l,+}^2\Sigma''_{l,0+}-\Phi_{l,-}^2\Sigma''_{l,0-})\, ,
\end{equation}
which can simply be absorbed into a re-definition of $H_0$ and will thus be neglected in the sequel.

Since the Hamiltonian is hermitian, we have $R_{+-,aa}\equiv\tilde{R}^{\, \prime}_{aa}+i \tilde{R}^{\, \prime\prime}_{aa}= R_{-+,aa}^*$.
Hence, together with $\Pi_R=-\Pi_L=\frac{1}{2}\sin\phi$ one finds
\begin{eqnarray}
\tilde{R}_{aa}^{\, \prime}&=&\frac{\sin\phi}{2}\sum_l(-1)^l(\Sigma_{-0l}'+\Sigma_{+0l}',-\Sigma_{0+l}',-\Sigma_{0-l}')\nonumber\\
\tilde{R}_{aa}^{\, \prime\prime}&=&\frac{\sin\phi}{2}\sum_l(-1)^l(\Sigma_{-0l}''-\Sigma_{+0l}'',\Sigma_{0+l}'',-\Sigma_{0-l}'')
\end{eqnarray}
which may have any signature, especially at non-zero bias $V_b=V_L-V_R\neq 0$ or for asymmetric couplings to
the leads $\Gamma_L\neq \Gamma_R$.  One {\em always} obtains an acceptable steady state for $V_b=0$.
This in turn motivates a perturbative treatment around the zero-bias situation (see second last section).

\subsection{The case of zero temperature}
The entries in the Redfield tensor simplify considerably in the zero temperature limit  $T\to 0$.  The self energy reduces to
\begin{eqnarray}
\Sigma^{'\rm in}_{l,ab}&=&\pi \Gamma_l\,  \Theta(-\omega_{l,ab})\nonumber\\
\Sigma^{''\rm in}_{l,ab}&=&-\Gamma_l\, (\gamma_E+\ln|\omega_{l,ab}/\omega_c|)\,
\end{eqnarray}
with the step function $\Theta(\omega)$. Accordingly, there are six generic cases with respect to the voltages
since we can always assume that $V_L>V_R$ and $e_+>e_-$, namely,

\begin{tabular}{c c c c|c}
$f_{L-0}\ \ $&$f_{L+0}\ \ $&$f_{R-0}\ \ $&$f_{R+0}\ \ $&$\ \rho_{\rm st}\ \ $\\
\hline
0&0&0&0&+\\
1&0&0&0&?\\
1&0&1&0&+\\
1&1&0&0&--\\
1&1&1&0&?\\
1&1&1&1&+\\
\end{tabular}\\

Here, we used for the Fermi functions the abbreviation $f_{lab}=f(\hbar \omega_{l, a b })=\Theta(-\omega_{l,ab})$. In the last column, cases marked with a '+' always yield acceptable steady states (positive populations), cases with a '--' yield
unphysical steady states (at least one population is negative) and cases with a '?' lead to unphysical steady states if $\Gamma_L\ne\Gamma_R$.

\section{Perturbative treatment for small bias}\label{perturbation}

For $V_b=0$ steady states are always acceptable stationary solutions. Hence, one may wonder if at least
a perturbative treatment of the the full Redfield approach (\ref{redfield1}) for low voltages always applies.
For this purpose, we expand the Redfield matrix around $V_b=0$: $R\approx  R_0+V_b\frac{\partial R}{\partial V_b}=R_0+V_b\delta R$.
The steady state
density is then given by $\rho_{\rm st}=\rho_0+V_b\delta\rho$ with $R_0\rho_0=0$. Note that $\rho_0$ in general differs from the density of the bare dot and thus accounts also for lead-dot hybridization as we discussed around (\ref{equi}).
The first order correction is determined from
\begin{equation}
R_0\, \delta\rho+\delta R\, \rho_0=0\,.\label{deltarho}
\end{equation}
As long as we are not near resonances such that $V_b$ differs sufficiently from transition energies $\hbar\omega_{ab}$, do
the columns of $\delta R$ lie in the space spanned by the columns of $R_0$. The
solution for $\delta\rho$ is then simply calculated.
Close to a resonance though, (\ref{deltarho}) has no solution meaning that
the Redfield scheme suffers from exactly the same difficulties as the non-equilibrium
Green's function formalism\cite{pride}.

The steady state current (\ref{ieq2}) away from resonances is easily obtained. We have
$\langle I_l\rangle=V_b G_l$ with $G=\delta\rho\Sigma_0+\rho_0\delta\Sigma$. Since $G$ is an observable and
$\rho_0$ is real, only the real parts of $\delta\Sigma$ contribute to $G$ (the imaginary parts are symmetrized out),
which, however, vanish away from resonances. In addition, as seen from (\ref{deltarho}), in this regime the main contributions
to $\delta R$ come from the imaginary part $R''$. Thus, $\delta\rho$ consists almost entirely of coherences
and the conductivity reads
\begin{eqnarray}
G&=&\frac{e}{\hbar^2}\sum_{abc}\delta\rho_{ab}
\left[\Phi_{bc}\Phi_{ca}^\dagger\Sigma^{\rm in}_{abl}
-\Phi_{bc}^\dagger\Phi_{ca}\Sigma^{\rm out}_{abl}\right]\nonumber\\
&=&\frac{e}{\hbar^2}\sum_{abc}\delta\rho_{ab}
\left[\Phi_{bc}\Phi_{ca}^\dagger\Sigma(\hbar\omega_{ba})\right.\nonumber\\
&&\left.\hspace{1cm}-\Phi_{bc}^\dagger\Phi_{ca}\Sigma(-\hbar\omega_{ba})\right]\, ,
\end{eqnarray}
where $\Sigma(\hbar\omega_{ba})=\left.\Sigma^{\rm in}_{l, ba}\right|_{V_b=0}$. The conductivity is of order $\Gamma_{L/R}$ in the lead-dot coupling and thus captures sequential charge transfer only but keeps lead induced coherences between the dot states. We recall that $\delta \rho$ is determined from a physical state $\rho_0$ via (\ref{deltarho}) and not from the density of the uncoupled dot.

\section{Conclusions}\label{conclusions}
We are now in position to summarize the findings of the previous sections and to draw conclusions. These refer to general situations and may not necessarily apply to specific set-ups.

(i) The Redfield approach provides always physical steady states if  imaginary parts of the Redfield tensor, i.e.\ imaginary parts of the self-energies (\ref{selfenergy}), are neglected. However, since in general real and imaginary parts are  on the same order of magnitude, this procedure (even if appealing for practical purposes) is not justified. (ii) In steady state the Markov approximation always applies independent of the couplings $\Gamma_R, \Gamma_L$. (iii) At zero voltage, the steady state cannot be taken as a product state consisting of the bare dot and the bare lead densities, but must be calculated from the full Redfield equation which then always provides physical states. (iv)
For finite bias voltages,  the only consistent use of the Redfield formulation for steady states is to apply an additional perturbative expansion around zero bias voltage in the interaction representation; this procedure basically coincides with a linear response treatment. The correct expansion parameter for the formulation is then not just given by the coupling constants $\Gamma_L, \Gamma_R$.
(v) In the high temperature limit $\hbar\beta\omega_{l, ab}\ll 1$ with $\omega_{l, ab}\neq 0$,  the treatment is justified if $\beta (\Gamma_L+\Gamma_R)\ll 1$. Even in this regime must a finite bandwidth $\omega_c$ be kept  in (\ref{selfenergy}) to avoid divergencies in the imaginary parts of the self-energies.
(vi) In the low temperature domain, the low-voltage expansion applies if
\begin{equation}
|\hbar\omega_{ab}| \gg k_{\rm B} T, V_b, |\Gamma_L-\Gamma_R|, (\Gamma_L+\Gamma_R)\,
\end{equation}
with the net bias voltage $V_b=|V_R-V_L|$. Here, the inequalities result from the conditions for $V_l \left.\partial \Sigma_l/\partial V_l\right|_{V_l=0}\sim V_l\Gamma_l \left.\partial \Psi(1/2-i \omega_{l, ab}/\nu_1)/\partial V_l\right|_{V_l=0}$ to be sufficiently small. In particular, the Redfield approach is thus constraint to
voltages away from resonances and to  weaker couplings with limited asymmetry. Outside this domain, numerical solutions of the full Redfield equations  (\ref{redfield1}) lead to unphysical steady states: The numerics includes all higher order voltage contributions contained in  the second order ($\sim \Gamma^2$) Born-Markov expansion while corresponding voltage contributions contained in fourth and higher order ($\sim \Gamma^n, n\geq 4$) are neglected. This is inconsistent and can then also not be cured by going beyond  the conventional $\Gamma^2$ expansion.

The physical interpretation is that no matter how small the coupling to the leads may be, in the asymptotic long time limit where the steady state exists, correlations between agent (dot) and its environment (leads) may drive the compound far away from its bare structure.
This can happen either when external energy supplied by a voltage source induces resonances or when large asymmetries in the agent-environment coupling induce strong quantum fluctuations and coherences between energy eigenstates of the bare agent. The Redfield approach is only consistently be applicable in domains of parameter space where these processes are suppressed.

\section*{Acknowledgements}
We thank M. Grifoni, C. Timm, M. Thoss for valuable discussions. Financial support was provided by the DFG through SFB569 and the German-Israeli Foundation.

\end{document}